\documentclass[prl,twocolumn,showpacs,preprintnumbers,amsmath,amssymb,tightenlines,epsfig]{revtex4}

\usepackage{graphicx}
\usepackage{dcolumn}
\usepackage{bm}
\usepackage{epsf}

\begin{document}

\preprint{Version: Submitted, \today}
\title{Stability of continuously pumped atom lasers}
\author{S. A. Haine, J. J. Hope, N. P. Robins and C. M. Savage}
\affiliation{Department of Physics and Theoretical Physics, 
Australian National University,
ACT 0200, Australia}

\email{joseph.hope@anu.edu.au}

\begin{abstract}
A multimode model of a continuously pumped atom laser is shown to be unstable below a critical value of the scattering length.  Above the critical scattering length, the atom laser reaches a steady state, the stability of which increases with pumping.  Below this limit the laser does not reach a steady state.  This instability results from the competition between gain and loss for the excited states of the lasing mode.  It will determine a fundamental limit for the linewidth of an atom laser beam.
\end{abstract}

\maketitle


\textit{Introduction.}--The physics of lasers, both fundamental and applied, has been an important theme in modern physics. Recently, the development of dilute gas Bose-Einstein condensation \cite{BECreview} has opened up the study of \textit{atom} lasers: the matter wave analogues of optical lasers \cite{atomlaserreview}. Many of the defining characteristics of optical lasers, such as high spectral density, recur for atom lasers.  In this letter we report on a critical, and somewhat counter-intuitive, difference between atom and optical lasers. One might expect that an atom laser would behave most like an ideal optical laser when the two-body interactions between atoms were minimal, because this is true for photons in an optical laser. We show that this is not the case, and that stable atom laser operation requires sufficiently repulsive two-body interactions between the atoms.  That these interactions are necessary is particularly significant because they induce a phase diffusion in the lasing mode, which places a fundamental limit on the linewidth of the atom laser beam.

Outcoupling from magnetically trapped BEC has been achieved by using radio 
frequency radiation \cite{Bloch continuous,rfexpt} or Raman transitions \cite{Ramanexpt} to change the internal state of the atoms, causing them to be either untrapped or anti-trapped.  Interference has been observed in the output field \cite{ALInterference}, and quasi-continuous beams have been produced \cite{Bloch continuous, Ramanexpt}.  Coupling the atoms out more slowly causes the output beam to have a narrower linewidth, but this comes at the expense of the beam flux \cite{earlyJoeOutput}.  High spectral flux in optical lasers is generated through a competition between a depletable pumping mechanism and the damping of the lasing mode.  This means that a higher pumping rate leads not only to greater flux, but also to a narrower output spectrum \cite{gainnarrowing}.  An atom laser with gain-narrowing must also have a saturable pumping mechanism that operates at the same time as the damping \cite{definingAL}.

Atom lasers have been described semiclassically with a large number of modes, and with a full quantum mechanical model containing a few modes.  Multimode models, using the Gross-Pitaevskii equation \cite{BECreview}, are capable of including the spatial effects of the atomic field and the interatomic interactions \cite{NickAL}.  This makes them useful for describing non-pumped atom laser experiments in which these effects are dominant.  Multimode models are required to determine whether pumped lasers approach single mode operation.  As for optical lasers, if the lasing system can be accurately described as a single mode then the quantum statistics of that mode control the linewidth of the output beam.  This means that although a semiclassical model is useful to determine the stability of an atom laser, a full quantum mechanical model must be used to calculate the linewidth of the beam coming from a pumped atom laser.  Single (or few) mode quantum theories have been used to describe continuously pumped atom lasers \cite{SMAL}.  Unfortunately, these models cannot describe the multimode behaviour of the atom laser such as the non-Markovian nature of the output coupling \cite{nonMarkov}, except under the approximation that the lasing mode itself is single mode \cite{SSNMAL}.  It is also extremely difficult to include the effects of the atomic interactions in these models.  Atom laser linewidths will be limited by these interactions due to Kerr-like dephasing of the lasing mode \cite{ALfeedback}, as well as by thermal effects \cite{Graham}.

Including the spatial effects, interactions and quantum statistics of the atom field requires a quantum field theory calculation which may become tractable by using stochastic field theories derived from phase space methods \cite{stochasticAF}.  These stochastic field simulations reduce to the semiclassical model in the limit of zero noise, so the semiclassical stability of the relevant system must first be examined.  We describe a semiclassical, multimode model of an atom laser and investigate the effects of varying the strength of the atom-atom interactions relative to the trap strength.  This parameter 
can be controlled experimentally by using Feshbach resonances \cite{RbFeshbach} or by adjusting the trap parameters.  We find that the atom laser is stable only when the repulsive interactions between the lasing mode atoms are sufficiently strong. For a given pumping rate and trap parameters, this results in the model being stable above a critical scattering length and unstable otherwise.


\textit{Atom laser model.}--Although a full quantum field theory will be required to calculate the linewidth of an atom laser with interactions, the stability of such a device can be determined with a multimode semiclassical treatment.  We model the atom laser as a two-component semiclassical atomic field coupled to an incoherent reservoir of atoms described by a density $\rho({\bf x})$.  The first field, $\psi_{t}({\bf x})$, is trapped in a harmonic potential and will form the lasing mode.  The second field, $\psi_{u}({\bf x})$, is untrapped and will form the laser beam.  The coupling can be achieved either by radio frequency transitions between the electronic states \cite{Bloch continuous,rfexpt}, or by a Raman transition \cite{Ramanexpt}.  The fields obey equations of the Gross-Pitaevskii form, with damping and reservoir coupling. The dynamical equations for the coupled system are
\begin{eqnarray}
\nonumber
i \frac{d\psi_t}{dt} &&= (-\frac{\hbar}{2M}\nabla^2 
+\frac{V}{\hbar} 
-i\gamma_t^{(1)}+\frac{U_{tu}}{\hbar}|\psi_u|^2+\frac{U_{tt}}{\hbar}| 
\psi_t|^2 \\ \nonumber &&
-i\gamma_t^{(2)}|\psi_t|^2-i\gamma_{tu}^{(2)}|\psi_u|^2 +\frac{i}{2} 
\kappa_p \rho)\psi_t +\kappa_R e^{i{\bf k}.{\bf x}}\psi_u ,
\\
\nonumber
i\frac{d\psi_u}{dt} &&= 
(-\frac{\hbar}{2M}\nabla^2+\frac{Mgx}{\hbar} 
+\frac{U_{tu}}{\hbar}|\psi_t|^2 +\frac{U_{tt}}{\hbar}|\psi_u|^2 \\ 
\nonumber && -i\gamma_u^{(2)}|\psi_u|^2 
-i\gamma_{tu}^{(2)}|\psi_t|^2) \psi_u + \kappa_R e^{-i {\bf k}.{\bf 
x}} \psi_t , 
\\
\frac{d\rho}{dt} &&= r -\gamma_p\rho  -\kappa_p 
|\psi_t|^2 \rho + \lambda \nabla^2 \rho ,
\label{eq:eom}
\end{eqnarray}
where $M$ is the mass of the atom, $V$ is the trapping potential, $g$ is the acceleration due to gravity, which is assumed to be directed along the $x$ direction. $U_{ij}=4 \pi \hbar a_{ij}/M$ is the interatomic interaction between $\psi_i$ and $\psi_j$, $a_{ij}$ is 
the s-wave scattering length between those same fields, $\gamma_i^{(1)}$ is the loss rate of $\psi_i$ due to background gas collisions, $\gamma_i^{(2)}$ is the loss rate of $\psi_i$ due to two-body inelastic collisions between particles in that state, $\gamma_{tu}^{(2)}$ is the loss rate of each field due to two-body inelastic collisions between particles in the other electronic state, $\kappa_R$ is the coupling rate between the trapped field and the output beam, ${\bf k}$ is the momentum kick due to the coupling process, $\kappa_p$ is the coupling rate between the pump reservoir and the trapped field. For the pump reservoir, $r$ is the rate of density increase, $\gamma_p$ is the loss rate, and $\lambda$ is the spatial diffusion coefficient.

We have not included three-body losses, which are normally important. However, near a Feshbach resonance they may be negligible, as for $^{85}$Rb, and still allow a wide range for the scattering length \cite{RbFeshbach}.  Our results are not affected by this limitation, but since it is not known how to include three-body losses in stochastic field simulations, we ignore them to facilitate later extension to a quantum field model.

The pumping terms in the above equations are phenomenological, describing an irreversible pumping mechanism from a reservoir which can be depleted, but is replenished at a steady rate.  These two features are necessary for any pumping mechanism that generates gain-narrowing through the competition of the gain and loss processes of the lasing mode.

It has been shown that output coupling can induce a 
localised bound state \cite{nonMarkov,SSNMAL}.  When the atom laser is pumped, the population of this bound state increases indefinitely.  Gravity must be included in the model to make this state decay, and allow steady operation to be achieved.  For a given set of parameters, there is a maximum coupling rate $\kappa_R$ below which this metastable state has negligible effect on the total dynamics.  We have also included the inelastic losses, which improve the overall stability.


\textit{Stability of the atom laser model.}--Throughout the rest of this letter we analyse the model given in Eq.(\ref{eq:eom}) in one dimension, assuming that both the lasing mode and the output beam are strongly trapped in the transverse dimensions.  We use the atomic properties and loss rates for $^{85}$Rb near a Feshbach resonance, where the interatomic interactions can be tuned with the magnetic fields \cite{RbFeshbach}.  We will use the following parameters for all subsequent calculations:
$\gamma^{(1)}_{t} = 7.0\times 10^{-3}$ s$^{-1}$, $\gamma^{(2)}_{t} =1.7 
\times 10^{-8}$ ms$^{-1}$,  $\gamma^{(2)}_{u} = 3.3\times 10^{-9}$ ms$^{-1}$, $ \gamma^{(2)}_{tu} = 0.5\gamma^{(2)}_{t}$, $\gamma_{p} = 5$ s$^{-1}$, 
$\kappa_{p} = 6.3 \times 10^{-4}$ ms$^{-1}$, a trapping potential of $V=m\omega^2 x^2/2$, where $\omega = 50$ rad s$^{-1}$, $|\mathbf{k}| = 10^{6}$ m$^{-1}$, $\lambda = 0.01$ m$^2$s$^{-1}$.  The interatomic interactions between different Zeeman levels are unknown for $^{85}$Rb, so we assume: $U_{t} = U_{u} = 2 U_{tu}$, a choice which has negligible effect on the results of this work.

\begin{figure}
\setlength{\epsfxsize}{8.6cm}
\centerline{\epsfbox{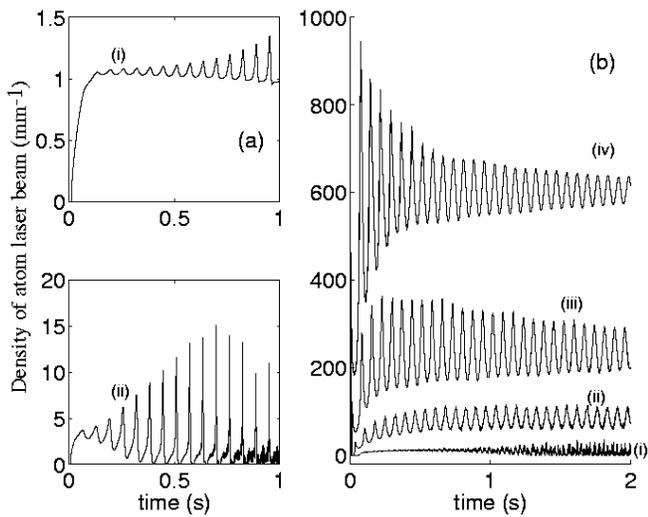}}
\caption{Density of the atom laser beam (in mm$^{-1}$), at a point below the trapped region, as a function of time.  (a) Scattering length $a_{tt} = 0$ and coupling rate $\kappa_R=10$ s$^{-1}$.  Pumping rates are (i) $R = 10^{5}$ s$^{-1}$ and (ii) $R = 10^{6} $ s$^{-1}$.  (b) $a_{tt} = 20$ nm and $\kappa_R=300$ s$^{-1}$.  Pumping rates are (i) $R = 500 $s$^{-1}$, (ii) $R = 10^{4} $ s$^{-1}$, (iii) $R = 10^{5} $ s$^{-1}$, and (iv) $R = 10^{6} $ s$^{-1}$.  All other parameters are as given in the text.}
\label{fig:AL}
\end{figure}

We integrate Eqs.(\ref{eq:eom}) numerically using a psuedo-spectral method with a Runge-Kutta time step \cite{RK4IP}.  It was found that the atom laser model exhibits varying levels of stability depending on both pumping rate and scattering length.  Figure \ref{fig:AL} shows the density of the atom laser beam at a point below the condensate for various pumping rates, and for two scattering lengths.  The total pumping rate is $R=r L$, where $L$ is the width of the pumping region.  We find that for zero scattering length the laser is unstable, with increasing pumping leading to modal oscillations.  In contrast, for large scattering length the atom laser becomes stable above a threshold pumping rate, and increasingly stable as the pumping rate increases.

The dominant oscillation of the laser mode is a breathing mode.  As we shall see, increasing the pumping increases the damping of this, and the higher frequency, modes, but only at a high scattering length. Varying the scattering length between the two cases shown in Fig.~1 gives intermediate types of behaviour, ranging from unstable to stable, with an increasing scattering length in the stable case making the oscillations more strongly damped. 


\textit{Analysis of the $a_{tt}=0$ case.}--With two-body interactions off, the spatial shape of the lasing mode is independent of the number of atoms in the trap.  An atom laser operating in this regime experiences no nonlinear phase diffusion, but then our atom laser model is not stable.  The semiclassical instability of this system can be understood with a simplified model, as it is directly related to the eigenfunctions of the lasing mode.  In this model we ignore the output coupling and hence the equation for the laser mode becomes
\begin{equation}
i \frac{d\psi_{t}}{dt} = (\frac{\hat{T} + V}{\hbar}  - i \gamma_{t}^{(1)}
- i \gamma_{t}^{(2)}|\psi_{t}|^{2} + 
\frac{i }{2}\kappa_p N_{p} ) \psi_{t} ,
\label{GPnobeam}
\end{equation}
where $\hat{T}$ is the kinetic energy operator, and $N_{p}= \int^{\infty}_{-\infty}\rho dx$.  Expanding Eq.(\ref{GPnobeam}) into the basis of harmonic oscillator eigenfunctions $\{\phi_n(x): (\hat{T} + V)\phi_n = E_n \phi_n\}$ and solving for the mode coefficients gives the equivalent set of equations
\begin{eqnarray}
\dot{c_{n}} = -\frac{i}{\hbar}E_{n}c_{n} + \frac{\kappa_p N_{p}}{2} c_{n} 
-\gamma_{t}^{(1)} c_{n}
-\gamma_{t}^{(2)}\sum_{ijk} c_{i}^{*}c_{j}c_{k}I_{nijk} 
\nonumber
\end{eqnarray}
with $I_{nijk}=\int_{-\infty}^{\infty}\phi_{n}^{*}\phi_{i}^{*}\phi_{j}\phi_{k}dx$ and 
$E_{n} = (n + \frac{1}{2})\hbar\omega$, and where 
we interpret $|c_{n}|^{2}$ as the number of atoms in mode $n$. The two-body 
loss term serves as a form of coupling between modes of equal parity, with 
the diagonal terms $I_n=I_{nnnn}$ acting as damping for each mode. The 
first few diagonal terms are $I_{0}= 0.5\beta, I_{1} = 0.375\beta, I_{2}= 0.320\beta, I_{3}= 0.287\beta, I_{4} = 0.264\beta, I_{5}= 0.244\beta$ with $\beta = \sqrt(\frac{2 m \omega}{\pi \hbar})$.  They decrease with increasing $n$, so the excited states experience lower loss than the lower energy states.  The loss rate decreases because the excited states are more spread out than the lower energy states.  This same trend occurs for three-body loss.  The pumping is not mode selective, so the excited states have a higher net gain.  The coupling between modes seeds excited states which become more populated than lower states.  This growth is the origin of the atom laser instability. 
%
%

\textit{Stability of a pumped Bose-Einstein condensate.}--With two-body interactions on, the previous simplified model is inadequate.  
As the interactions between the atoms become dominant, the excited states have a different spatial shape, and may become more damped than the ground state.  For mode-insensitive pumping this would lead to a selection of the ground state.

Hence we expect there to be a critical scattering length for which the atom laser moves from unstable to stable. Since the nonlinear interactions depend on the density, we also expect the critical scattering length to be dependent on the pump rate and trapping frequency.  

We found that the semiclassical dynamics of the atom laser beam followed the dynamics of the lasing mode condensate. For the range of output coupling rates that we consider, the condensate dynamics is only weakly affected by the output coupled beam. Hence, in order to reduce the computational load, we ignored the output coupling in the following.
\begin{figure}
\setlength{\epsfxsize}{8.6cm}
\centerline{\epsfbox{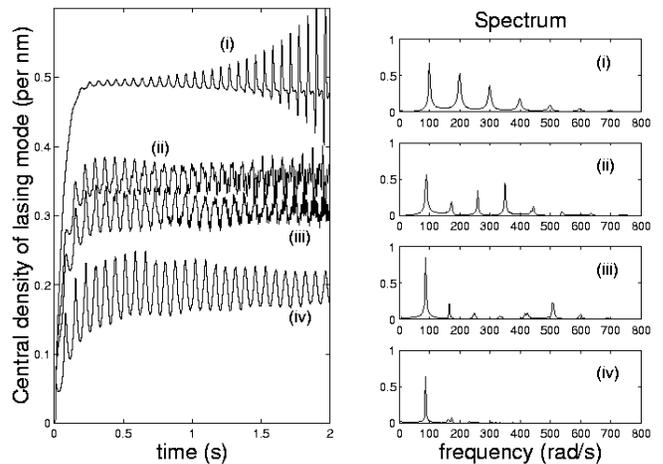}}
\caption{\label{fig:scattering} Time series and power spectra of the laser mode central density.  The pumping rate is $R = 10^{5}$ s$^{-1}$.  The scattering lengths are (i) $a_{tt} = 0.01$ nm, (ii) $a_{tt}= 0.5$ nm, (iii) $a_{tt} = 1$ nm, and (iv) $a_{tt}= 10$ nm.   Spectral power is in arbitrary units and is obtained from the time series after the initial $0.4$ s.}
\end{figure}

Figure \ref{fig:scattering} shows the different regimes of stability as the scattering length is changed with fixed pumping rate.  The case with the lowest scattering length is unstable due to the excitation of higher trap modes.  The system with the highest scattering length is stable, with all excited states being damped.  The systems with intermediate scattering length show a reasonably steady population in the lowest excited states, and a slow growth of high frequencies due to the population of the excited states. 
Figure \ref{fig:scattering} also shows the power spectrum of the oscillations in the central density, ignoring the transients in the first $0.4$ seconds of each trace.  The spectrum of the case with the lowest scattering length shows that the behaviour resembles that of the harmonic oscillator, with growing oscillations at the frequencies $2n\omega=100 n/2\pi$ Hz, which are the even harmonic oscillator eigenfrequencies.  At larger scattering lengths, the behaviour is stable, but the transient oscillations closely resemble the even Thomas-Fermi eigenfrequencies $\omega \sqrt{n(n+1)/2}$ ($n$ even) \cite{NickAL,Kneer}. 

\begin{figure}
\setlength{\epsfxsize}{8.6cm}
\centerline{\epsfbox{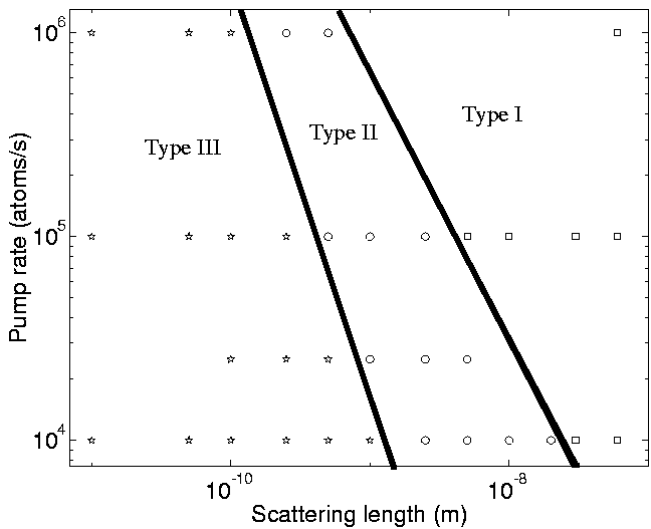}}
\caption{\label{fig:phasespace} Atom laser stability in pumping rate and scattering length parameter space (log-log scales).  Cases which we have determined to be in the stable region (Type I) are shown as squares, Type II systems are shown as circles, and unstable (Type III) systems are shown as stars.  The bold lines separate the stability regions.}
\end{figure}

We found it convenient to classify the behaviour into three catogories. Type I behaviour is where all excited modes are damped.  In Type II behaviour the breathing mode is damped but higher modes are undamped, so oscillations are initially decreasing, but the system is unstable over a long period.  In Type III behaviour all modes are undamped, and the system is unstable.  Figure \ref{fig:phasespace} shows the type of behaviour found for different values of the pumping rate $R$ and scattering length $a_{tt}$.  For high pumping rates, the high frequency oscillations produce numerical instabilities, so the boundaries in the phase diagram were difficult to explore in detail. 

The semiclassical stability of the pumped condensate is almost purely determined by the pumping rate and the scattering length, as these two parameters control the mean field interaction strength.


 \textit{Conclusions:}  We have shown that a semiclassical atom laser becomes stable above a critical scattering length.  This is because the excited states are increasingly damped when the interactions are the dominant contribution to the lasing mode energy, and otherwise they are more weakly damped than the lower energy lasing modes.  This effect is not dependent on the output coupling, but could possibly be controlled by a highly mode selective pumping mechanism.  This result is important in the design of continuously pumped atom lasers because the Kerr-like phase diffusion will scale with the interactions, and will induce a fundamental limit to the linewidth of the atom laser beam \cite{ALfeedback}.  Due to the opposite scaling of the semiclassical and the quantum noise, if an atom laser can be operated in a stable regime then we expect it to have a minimum linewidth at an optimum scattering length.

\begin{acknowledgments}
This research was supported by the Australian Research Council and 
by an award under the Merit Allocation Scheme on the National Facility of the Australian Partnership for Advanced Computing. 
\end{acknowledgments}

\end{document}